\documentclass[preprint,3p,twocolumn,number,sort&compress]{elsarticle}

\usepackage{refcount}                      
\usepackage{fmtcount}                      
\usepackage{amsmath}                       
\usepackage{amssymb}                       
\usepackage{booktabs}                      
\usepackage{subcaption}                    
\usepackage[inline]{enumitem}              
\usepackage[pdftex]{hyperref}              
\usepackage{color}                         

\makeatletter
\newcommand{\pushright}[1]{\ifmeasuring@#1\else\omit\hfill$\displaystyle#1$\fi\ignorespaces}
\newcommand{\pushleft}[1]{\ifmeasuring@#1\else\omit$\displaystyle#1$\hfill\fi\ignorespaces}
\makeatother

\newcommand{\up}[1]{\textsuperscript{#1}}  
\newcommand{\ie}[1]{\textit{i.e.}}         
\newcommand{\origin}{\textcolor{red}{$\bullet$}}

\begin{document}
\begin{frontmatter}
  \title{Numerical analysis of a folded superconducting coaxial shield
    for cryogenic current comparators}
  \author[temf]{Nicolas~Marsic\corref{cor1}}
  \ead{marsic@temf.tu-darmstadt.de}
  \author[temf]{Wolfgang~F.~O.~M\"uller}
  \author[temf]{Herbert~De~Gersem}
  \author[ipht]{Matthias~Schmelz}
  \author[ipht]{Vyacheslav~Zakosarenko}
  \author[ipht]{Ronny~Stolz}
  \author[gsi] {Febin~Kurian}
  \author[gsi] {Thomas~Sieber}
  \author[gsi] {Marcus~Schwickert}

  \cortext[cor1]{Corresponding author}
  \address[temf]{Institut f\"ur Theorie Elektromagnetischer Felder,
    Technische Universit\"at Darmstadt,
    Schlossgartenstra\ss{}e 8, 64289 Darmstadt, Germany}
  \address[ipht]{Leibniz Institute of Photonic Technology,
    Albert-Einstein-Stra\ss{}e 9, 07745 Jena, Germany}
  \address[gsi] {GSI Helmholtzzentrum f\"ur Schwerionenforschung,
    Planckstra\ss{}e 1, 64291 Darmstadt, Germany}
  \date{}

  \begin{abstract}
    This paper presents a new shield configuration
    for cryogenic current comparators (CCCs),
    namely the folded coaxial geometry.
    An analytical model describing its shielding performance is first developed,
    and then validated by means of finite element simulations.
    Thanks to this model,
    the fundamental properties of the new shield are highlighted.
    Additionally,
    this paper compares the volumetric performance
    of the folded coaxial shield to the one of a ring shield,
    the latter being installed in many CCCs for measuring particle beam currents
    in accelerator facilities.
  \end{abstract}

\end{frontmatter}

\section{Introduction}
\begin{figure*}[ht]
  \centering
  \subcaptionbox{Ring topology.
    \label{fig:schematic:ring}}
  {\vspace{1.17cm}\includegraphics[width=4.5cm]{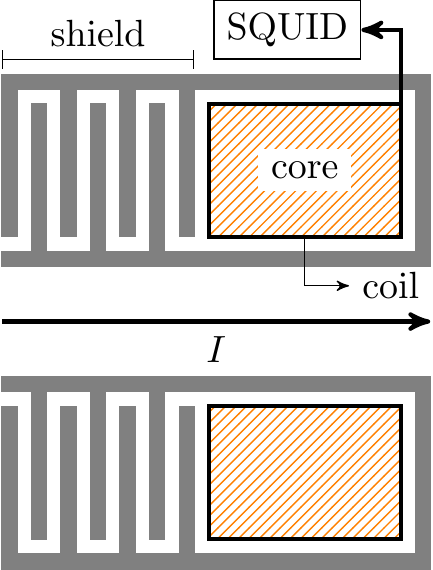}}
  \hfill
  \subcaptionbox{Folded coaxial topology (opening at the outer radius).
    \label{fig:schematic:dw}}
  {\includegraphics{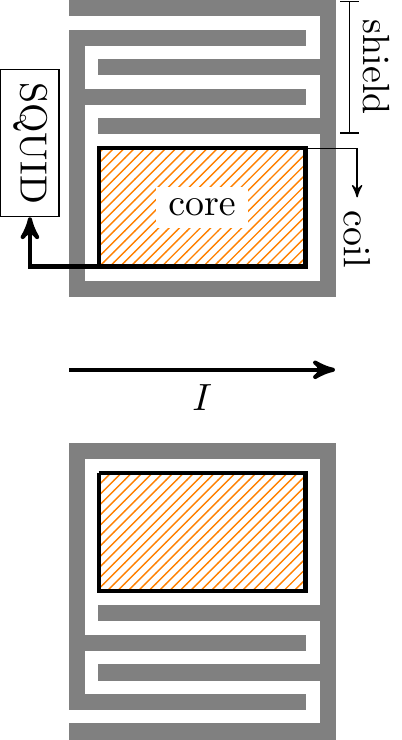}}
  \hfill
  \subcaptionbox{Folded coaxial topology (opening at the inner radius).
    \label{fig:schematic:up}}
  {\includegraphics{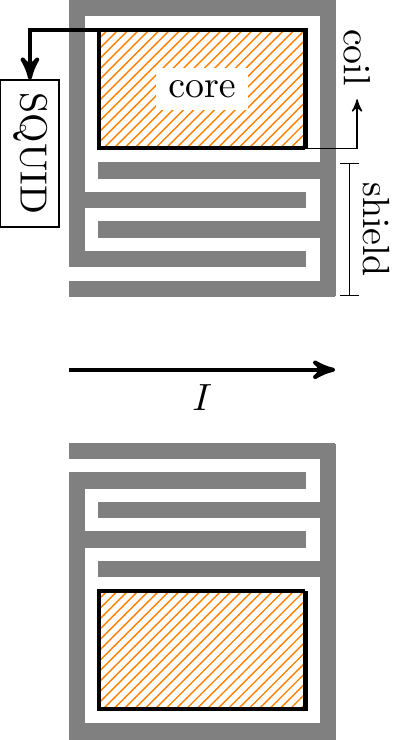}}
  \caption{Three possible cryogenic current comparators
      (axial cuts, two dimensional schematics).}
  \label{fig:schematic}
\end{figure*}

Nowadays, cryogenic current comparators (CCCs) are the most sensitive
instruments to measure very low electric currents with high accuracy.
A typical CCC consists of a superconducting shield
separating the current to be measured
and a zero magnetic flux detector~\cite{Harvey1972, Williams2011},
the latter usually being implemented
by a superconducting quantum interference device (SQUID)~\cite{Clarke2004}.
The magnetometer is then coupled with the system through a detection coil
and possibly a pickup core, as shown in Figure~\ref{fig:schematic} for different
geometrical and topological configurations.

The aim of the CCC shield is to reject any external magnetic flux,
while remaining transparent to the induction field originated
from the electric current to be measured.
This property is obtained by choosing an appropriate shield geometry.
Furthermore, the specific construction leads to current measurements which are
insensitive to the transverse displacement of the line current
with respect to the symmetry axis of the CCC.
These features are the keystones of the CCC's exceptional precision,
but are only achievable with superconducting materials.
Many shield topologies and geometries have been proposed and studied
in the literature~\cite{Sullivan1974, Grohmann1974,
  Grohmann1976a, Grohmann1976b, Seppa1990}.
In this work,
we investigate the performance of a folded coaxial shield
(depicted in Figures~\ref{fig:schematic:dw} and~~\ref{fig:schematic:up}),
which is, to the best of our knowledge, studied here for the first time.
This shield configuration is compared to the ring structure
(depicted in Figure~\ref{fig:schematic:ring})
found in several CCCs
dedicated to the diagnostics of particle beams in accelerator facilities~\cite{
  Peters1998, Tanabe1999, Geithner2012, Fernandes2017}.
In this comparison,
the shields geometries are constrained by the radius of the beam tube
and the cross-sectional area of the CCC pickup core.

This paper is organized as follows.
In section~\ref{sec:coax},
the shielding performance of the coaxial shield is reviewed.
Then, in section~\ref{sec:fold},
the new folded coaxial shield is introduced
and its performance is analyzed.
To this end, an analytical model is first developed,
and then validated by means of finite element simulations.
The paper continues by reviewing the shielding properties
of the ring configuration in section~\ref{sec:ring}.
A performance comparison between the ring and folded coaxial topologies is
then presented in section~\ref{sec:perf}.
Finally, conclusions are drawn in section~\ref{sec:conclusion}.

\section{Coaxial shield}
\label{sec:coax}
\begin{figure}[ht]
  \centering
  \includegraphics{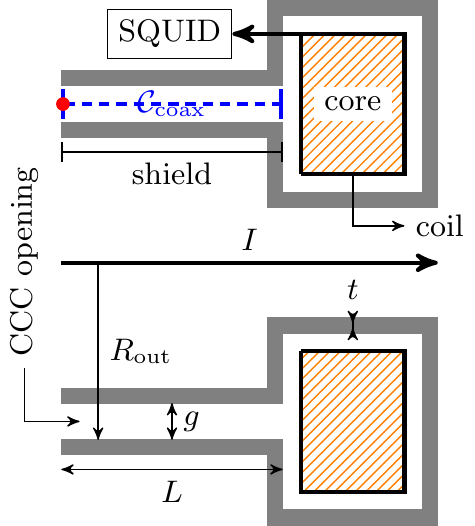}
  \caption{Coaxial configuration of a CCC shield
    (axial cut, two dimensional schematic);
    the symbol~\origin~indicates the origin of a curve
    (\ie{}~the location of its parametric coordinate $s_\text{min}$).}
  \label{fig:coax:2}
\end{figure}

\begin{figure}[ht]
  \centering
  \includegraphics[width=4.0cm]{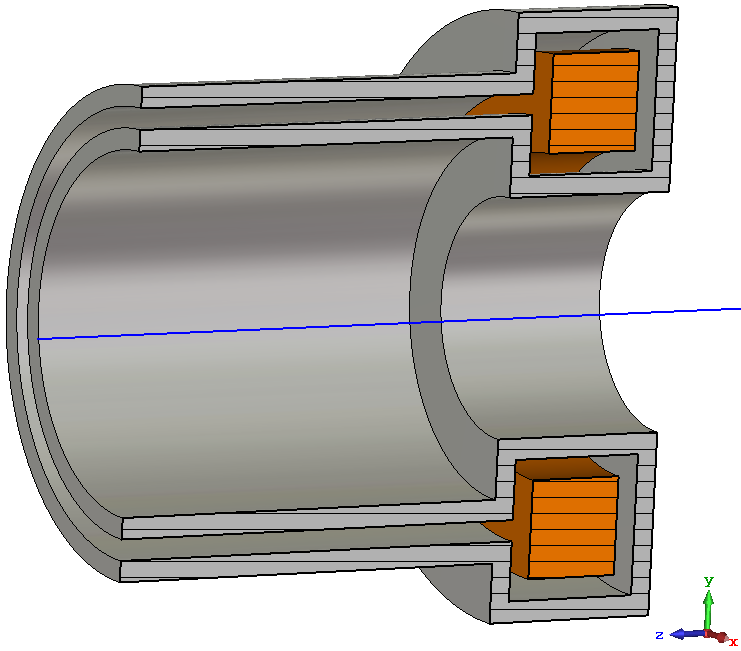}
  \caption{Coaxial configuration of a CCC shield
    (axial cut, three dimensional CAD view, CCC axis depicted as a blue line).}
  \label{fig:coax:3}
\end{figure}

\begin{figure}[ht]
  \centering
  \includegraphics[width=6.0cm]{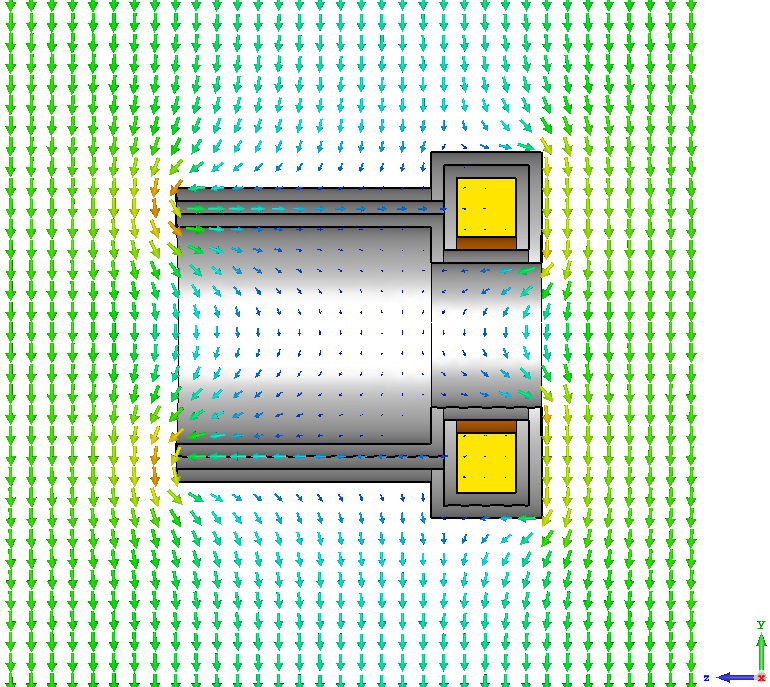}
  \caption{Example of a source magnetic induction field
    undergoing the weakest damping from the shield
    (axial cut, arbitrary scale).}
  \label{fig:coax:bDipole}
\end{figure}

Let us start by reviewing the classical results for the coaxial shield
presented in Figures~\ref{fig:coax:2} and~\ref{fig:coax:3}.
It can be shown that this shield configuration attenuates
every component of the magnetic induction field,
with the exception of the azimuthal one~\cite{Grohmann1976a}.
Moreover, the theory reveals that the component undergoing
the weakest (but still existing) damping from the shield has the following property:
it is spatially constant (in Cartesian coordinates) and perpendicular to the CCC axis,
as shown in Figure~\ref{fig:coax:bDipole}.
Thus, it is sufficient to analyze the damping experienced by this component
in order to assess the performance of the shield.
Let us formally define the \emph{global} damping profile $d_\mathcal{C}(s)$
and the \emph{local} damping profile $\delta_\mathcal{C}(s)$
of the \emph{least damped field} along a curve~$\mathcal{C}$
parameterized by $s\in[s_\text{min},s_\text{max}]\subset\mathbb{R}$:
\begin{subequations}
  \label{eq:damping}
  \begin{align}
    d_\mathcal{C}(s) = & \displaystyle\frac{B_\text{in}}{B(s)},
                         \label{eq:damping:glob}
    \\
    \delta_\mathcal{C}(s) = & \displaystyle\frac{B(s_\text{min})}{B(s)},
                              \label{eq:damping:loc}
  \end{align}
\end{subequations}
where $B$ is the magnitude of the magnetic induction and
$B_\text{in}$ the magnitude of the magnetic induction \emph{at the CCC opening}.

The only difference between these two definitions
is the normalization procedure:
\begin{enumerate*}[label = \itshape\roman*\upshape)]
\item in the global case, the damping is normalized with respect to
    the field intensity at the CCC opening;
\item in the local case, the damping is normalized with respect to
    the field intensity at the origin of $\mathcal{C}$,
    which depends on the definition of the curve.
\end{enumerate*}
Let us note that in the example given in Figure~\ref{fig:coax:2},
the quantities $d_{\mathcal{C}_\text{coax}}(s)$ and $\delta_{\mathcal{C}_\text{coax}}(s)$
are identical, since the origin $s_\text{min}$ of $\mathcal{C}_\text{coax}$
is located at the CCC opening.
On the other hand, if we take the example of the curve $\mathcal{C}^2_\text{folded}$
depicted in Figure~\ref{fig:folded}, the values given by
$d_{\mathcal{C}^2_\text{folded}}(s)$ and $\delta_{\mathcal{C}^2_\text{folded}}(s)$
will be different, since $s_\text{min}$ is not located at the CCC opening.
Of course, in order to assess the performance of a shield,
the global damping profile is of interest.
However,
the local variant eases the analysis of complex structures.

With this definition in hand,
the global damping profile of the coaxial shield is~\cite{Grohmann1976a}:
\begin{equation}
  \label{eq:coax:glob}
  d_{\mathcal{C}_\text{coax}}(s) =  \exp\left(\frac{s}{R_\text{out}}\right)
  \quad\forall s\in[0, L],
\end{equation}
where $R_\text{out}$ is the outer radius of the coaxial shield,
and where the curve $\mathcal{C}_\text{coax}$ is a straight segment
starting at the CCC opening
and traveling along the whole shield of length $L$,
as depicted in Figure~\ref{fig:coax:2}.
The coaxial structure thus offers an exponential damping profile.
For a particular section $\widetilde{\mathcal{C}}_\text{coax}$
of $\mathcal{C}_\text{coax}$, the local damping is given by:
\begin{equation}
  \label{eq:coax:loc}
  \delta_{\widetilde{\mathcal{C}}_\text{coax}}(s) =
  \exp\left(\frac{s-s_\text{min}}{R_\text{out}}\right)
  \,\,\,\forall s\in[s_\text{min}, s_\text{max}],
\end{equation}
where $0 \leq s_\text{min} \leq s_\text{max} \leq L$.
Finally, let us mention that the above results have been derived
under the following assumptions~\cite{Grohmann1976a}:
\begin{enumerate}
\item the superconducting material is assumed
  to be ideal and perfectly described by the London equations;

\item the thickness (denoted by~$t$ in Figure~\ref{fig:coax:2})
  of the superconducting shield is larger than its London penetration depth,
  so that perfect diamagnetism can be assumed;

\item the thickness of the superconducting shield,
  as well as the size of the shield air gap
  (denoted by~$g$ in Figure~\ref{fig:coax:2}),
  are small compared to the inner and outer radii of the CCC.
  \label{enum:hyp:t}
\end{enumerate}

\section{Folded coaxial shield}
\label{sec:fold}
\begin{figure}[ht]
  \centering
  \includegraphics{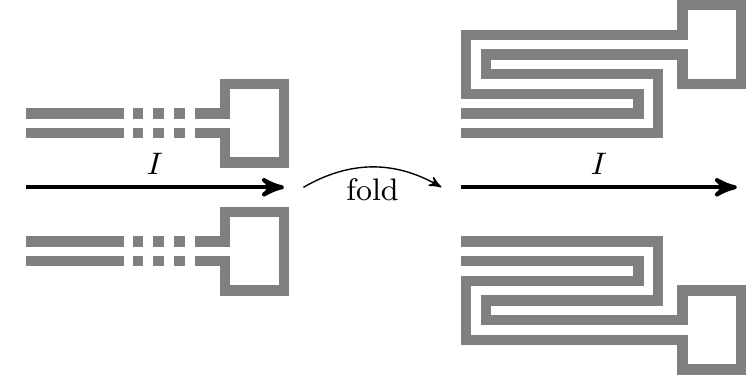}
  \caption{Folding a coaxial shield into a meander structure (axial cut).}
  \label{fig:folded:coax}
\end{figure}

From~\eqref{eq:coax:glob}, it is evident that
the overall attenuation achieved by the shield depends on its length.
Unfortunately, this can lead to a structure with an excessive size in the axial direction.
A natural solution to circumvent this limitation is to fold the coaxial structure
in the radial direction,
as indicated in Figure~\ref{fig:folded:coax}.
This leads to a folded coaxial shield forming a meander-like structure,
as depicted in Figure~\ref{fig:folded}.

\subsection{Two possible variants for the folded coaxial CCC}
\begin{figure*}[ht]
  \centering
  \subcaptionbox{Three dimensional CAD view of the outer variant
    (opening at the outer radius).}
  {\includegraphics[width=3.6cm]{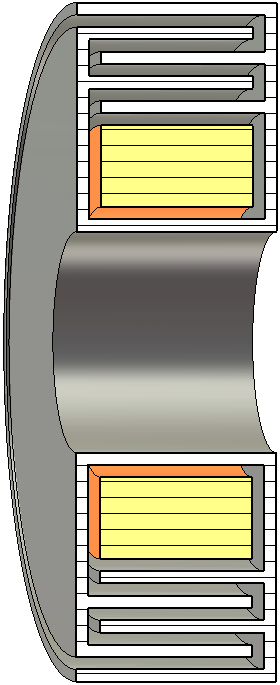}}
  \hfill
  \subcaptionbox{Two dimensional schematic of the outer variant
    (opening at the outer radius).}
  {\includegraphics{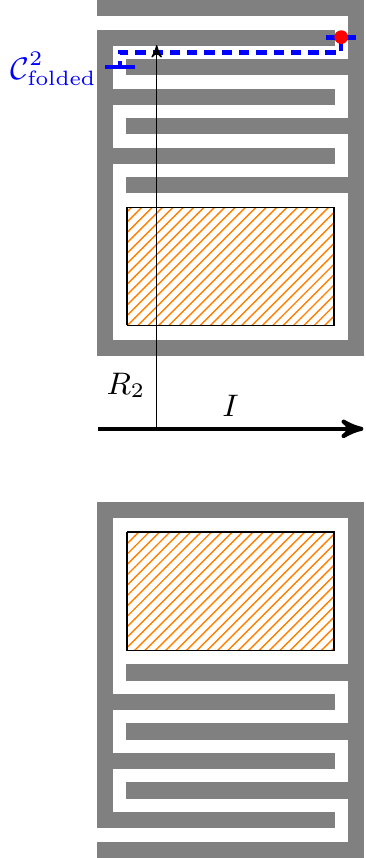}}
  \hfill
  \subcaptionbox{Two dimensional schematic of the inner variant
    (opening at the inner radius).}
  {\includegraphics{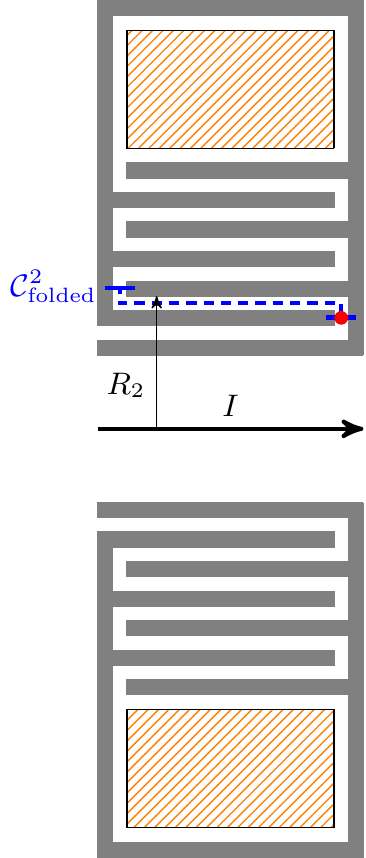}}
  \caption{The two possible variants for the folded coaxial CCC (axial cuts);
      the symbol~\origin~indicates the origin of a curve
      (\ie{}~the location of its parametric coordinate $s_\text{min}$).}
  \label{fig:folded}
\end{figure*}

After folding of the coaxial shield structure,
two possibilities remain for completing the CCC:
the detection coil is either located at the inner, or at the outer radius of the shield,
as shown in Figure~\ref{fig:folded}.
Evidently, the opening of the folded coaxial shield
is located at the opposite radius.
In the following,
the \textit{inner} variant refers to the CCC
with the \emph{meander opening} at the inner radius,
and the \textit{outer} variant refers to the other case.
As discussed later in this paper,
both variants show different behavior.

\subsection{Analytical damping profile}
Analyzing the shape of a folded coaxial shield,
we can directly notice that the structure is
a set of coaxial cylinders concentrically stacked around each others.
Therefore, for one shell of the stack,
the local damping defined in~\eqref{eq:coax:loc} holds,
if an appropriate redefinition of the curve $\mathcal{C}$
and its corresponding outer radius $R_\text{out}$ is given.

Let us start by considering a folded coaxial shield
consisting of a stack of $N$ shells,
each shell being identified by an index $i\in\{1, \dots, N\}$.
For the $i$\up{th} shell, the curve $\mathcal{C}_\text{folded}^i$ is composed
of the $i$\up{th} straight section and
half of the $(i-1)$\up{th} and $(i+1)$\up{th} linking parts,
as depicted in Figure~\ref{fig:folded}.
We then introduce a parameter $s\in[\ell_{i-1}, \ell_i]$
to parametrize $\mathcal{C}_\text{folded}^i$,
where $\ell_i$ is the total \emph{unfolded} length
between the \emph{CCC opening} and the end of $\mathcal{C}_\text{folded}^i$.
By convention, we further impose that $\ell_0 = 0$.
With these definitions, and by exploiting~\eqref{eq:coax:loc},
the \emph{local} damping introduced by the $i$\up{th} shell writes:
\begin{equation}
  \label{eq:folded:loc}
  \delta_{\mathcal{C}_\text{folded}^i} =
  \exp{\left(\frac{s-\ell_{i-1}}{R_i}\right)}
  \quad\forall s\in[\ell_{i-1}, \ell_i],
\end{equation}
where $R_i$ is the outer radius of the $i$\up{th} shell.
It is worth mentioning
that this equation does not consider the variation of $R_i$
along the linking parts constituting $\mathcal{C}_\text{folded}^i$.
However, thanks to the \ordinalstringnum{\getrefnumber{enum:hyp:t}} hypothesis
(see section~\ref{sec:coax}), this simplification is legitimate.

The \emph{global} damping $d_{\mathcal{C}_\text{folded}^i}$
introduced by the $i$\up{th} shell is computed as follows.
By combining~\eqref{eq:damping:loc} and~\eqref{eq:damping:glob},
we have:
\begin{equation}
  \label{eq:damping:mix}
  d_{\mathcal{C}}(s) =
  \frac{B_\text{in}}{B(s_\text{min})}\delta_{\mathcal{C}}(s)
  \quad\forall s\in[s_\text{min}, s_\text{max}].
\end{equation}
Thus, by inserting~\eqref{eq:folded:loc} in~\eqref{eq:damping:mix},
we can write:
\begin{equation}
  \label{eq:folded:glob}
  d_{\mathcal{C}_\text{folded}^i}(s) =
  \alpha_i \delta_{\mathcal{C}_\text{folded}^i}(s)
  \quad\forall s\in[\ell_{i-1}, \ell_i],
\end{equation}
where, by exploiting~\eqref{eq:damping:glob}
(the definition of $d_\mathcal{C}$), the coefficient $\alpha_i$ is given by
\begin{equation}
  \label{eq:folded:a:def}
  \alpha_i = \frac{B_\text{in}}{B(s_\text{min})} =
  d_{\mathcal{C}_\text{folded}^i}(\ell_{i-1}).
\end{equation}
Physically speaking,
this coefficient $\alpha_i$ accounts for the damping
introduced by the $(i-1)$ previous shells.

Let us now show that the following holds:
\begin{equation}
  \label{eq:folded:a}
  \alpha_1=1
  \quad
  \text{and}
  \quad
  \alpha_{i+1}=\alpha_i\exp{\left(\frac{\ell_i-\ell_{i-1}}{R_i}\right)}.
\end{equation}
Since for the first shell we have $B(s_\text{min}) = B_\text{in}$
by definition of $\mathcal{C}_\text{folded}^1$,
we can directly exploit~\eqref{eq:folded:a:def} to show that $\alpha_1=1$.
Furthermore,
by combining the definitions~\eqref{eq:folded:glob}, \eqref{eq:folded:a:def}
and the local damping~\eqref{eq:folded:loc}, we can write:
\begin{align}
  d_{\mathcal{C}_\text{folded}^i}(s) =
  d_{\mathcal{C}_\text{folded}^i}(\ell_{i-1})
  \exp{\left(\frac{s-\ell_{i-1}}{R_i}\right)}\nonumber\\
  \forall s\in[\ell_{i-1}, \ell_i].
  \label{eq:folded:induction:i}
\end{align}
Moreover, because of the continuity of the magnetic induction, 
we have that:
\begin{equation}
  \label{eq:folded:continuity}
  d_{\mathcal{C}_\text{folded}^{i+1}}(\ell_{i}) =
  d_{\mathcal{C}_\text{folded}^{i}}(\ell_{i}).
\end{equation}
Thus, by evaluating \eqref{eq:folded:induction:i} at $s=\ell_i$
and by using~\eqref{eq:folded:continuity}, the following holds:
\begin{align}
  \label{eq:folded:induction:i+1}
  d_{\mathcal{C}_\text{folded}^i}(\ell_i)
  & =
    d_{\mathcal{C}_\text{folded}^{i+1}}(\ell_i),\nonumber\\
  & =
    d_{\mathcal{C}_\text{folded}^i}(\ell_{i-1})
    \exp{\left(\frac{\ell_i-\ell_{i-1}}{R_i}\right)}.
\end{align}
Then, by exploiting once more the definition~\eqref{eq:folded:a:def},
we can conclude the proof:
\begin{equation}
  \label{eq:folded:induction:proof}
  \alpha_{i+1} = \alpha_i
  \exp{\left(\frac{\ell_i-\ell_{i-1}}{R_i}\right)}.
\end{equation}

From these last results,
it worth noticing that, the total damping exhibited by the shield
\begin{align}
  d_{\mathcal{C}_\text{folded}^{N}}(\ell_N)
  & = \alpha_N\delta_{\mathcal{C}_\text{folded}^{N}}(\ell_N),
    \nonumber\\
  & = \alpha_{N-1}
    \exp\left(\frac{\ell_{N-1}-\ell_{N-2}}{R_{N-1}}\right)\nonumber\\
  & \pushright{\exp\left(\frac{\ell_N-\ell_{N-1}}{R_N}\right),}
    \nonumber\\
  & = \dots,
    \nonumber\\
  & = \alpha_{N-(N-1)}\nonumber\\
  & \pushright{
      \phantom{=}~\exp\left(
        \frac{\ell_{N-(N-1)}-\ell_{N-(N-1)-1}}{R_{N-(N-1)}}
      \right)\dots
    }
    \nonumber\\
  & \pushright{
      \exp\left(\frac{\ell_N-\ell_{N-1}}{R_N}\right),
    }
    \nonumber\\
  & = \prod_{i=1}^N\exp\left(\frac{\ell_i-\ell_{i-1}}{R_i}\right),
\end{align}
appears naturally as the product of the damping introduced by each layer.

\subsection{Numerical validation}
To validate the analytical damping model~\eqref{eq:folded:glob}
and~\eqref{eq:folded:a},
we arranged a numerical finite element (FE) simulation,
utilizing the magnetostatic solver provided in CST EM STUDIO\up{\copyright}\cite{Cst2015}.
The superconducting parts of the CCC are considered
as perfect electrical conductors,
the relative magnetic permeability of the pickup core is set to
$\mu_r^\text{core} = 1000$
and the dimensions of the CCC are given in Table~\ref{tab:folded:fem}.
The simulation model embeds the CCC in a spatially constant magnetic field perpendicular to the CCC axis,
as the one depicted in Figure~\ref{fig:coax:bDipole}.
The computed magnetic induction field is then sampled along a path
following the meanders of the structure.
By normalizing the magnitude of the magnetic induction,
the damping introduced by the folded coaxial structure was assessed.
The numerical results were validated by comparing simulations with different
mesh curvatures and FE discretization orders.
In the following, the computed dampings are expressed in decibel (dB),
following the convention:
\begin{equation}
  \label{eq:db}
  d^\text{dB}_\mathcal{C}(s) = 20\log_{10}{\Big(d_\mathcal{C}(s)\Big)}.
\end{equation}
\begin{table}[ht]
  \centering
  \begin{tabular}{cccccc}
    \hline
    \hline
    Gap     & Thickness & Inner radius  \\
    \cmidrule(lr){1-1}\cmidrule(lr){2-2}\cmidrule(lr){3-3}
    $0.5$mm & $3$mm     & $120$mm       \\
    \\
    Axial length & Core height & Shells \\
    \cmidrule(lr){1-1}\cmidrule(lr){2-2}\cmidrule(lr){3-3}
    $120$mm      & $50$mm      & $24$   \\
    \hline
    \hline
  \end{tabular}
  \caption{Main geometrical parameters for the FE simulations.}
  \label{tab:folded:fem}
\end{table}

\begin{figure}[ht]
  \centering
  \includegraphics{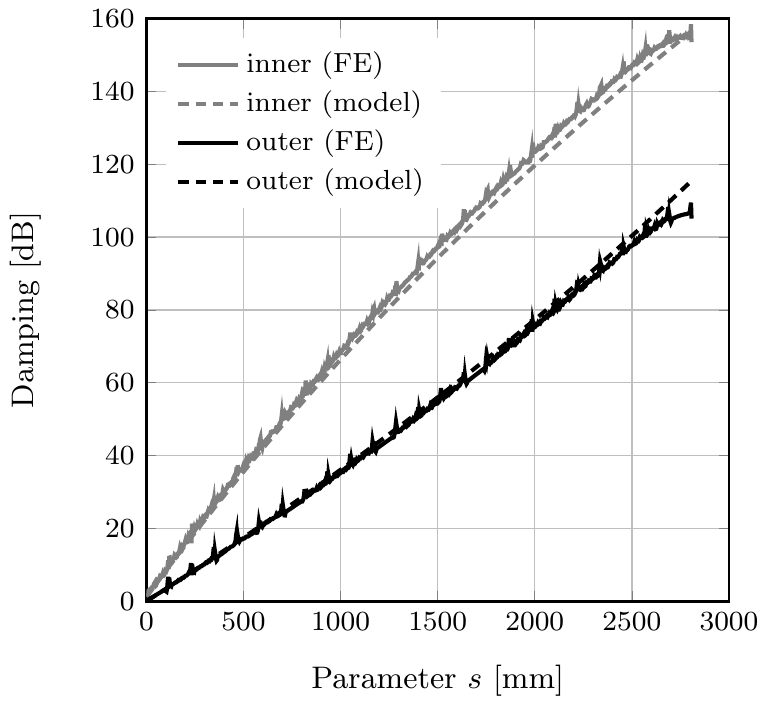}
  \caption{Comparison between FE simulations
    and equations~\eqref{eq:folded:glob} and~\eqref{eq:folded:a}.}
  \label{fig:folded:fem}
\end{figure}

The finite element results,
as well as the predictions given by the analytical model~\eqref{eq:folded:glob}
and~\eqref{eq:folded:a}, are summarized in Figure~\ref{fig:folded:fem}
for the inner and outer variants.
First, it can directly be noticed that the numerical solutions
and the analytical model match well,
which comforts us in the validity of our previous developments and assumptions.
Maximum relative deviations of $10\%$ and  $5\%$ are observed for the outer and inner variants respectively.
Secondly,
the variant with the meander opening at the inner radius
exhibits a larger final damping than the other variant.
This phenomenon is easily explained by the fact that the two variants
are not symmetric.
Indeed, because of the detection coil, the radii of the coaxial shells
forming the shield do not span the same range:
$R_i^\text{inner}\in[120, 201]$mm for the inner variant and
$R_i^\text{outer}\in[254, 173]$mm for the outer one.
Furthermore, and as explained in the previous subsection, the local damping
introduced by the $i$\up{th} coaxial shell is given by~\eqref{eq:folded:loc}
and is proportional to $\exp(1/R_i)$.
Therefore, since the values taken by the sequence $R^\text{inner}_i$
are smaller or equal to the values taken by the sequence $R^\text{outer}_i$,
the inner variant must exhibit a larger final damping.
Finally, we can observe that the damping profile of the inner variant
shows a decreasing derivative,
while the outer alternative shows an increasing one.
This behavior is again easily explained.
For the inner variant, as we enter deeper in the meander structure,
the damping decreases since the radius increases.
For the outer variant, the damping increases since the radius decreases.

To conclude our numerical validation,
we analyzed the influence of the air gap size
and of the superconducting material thickness on the damping
in more detail.
Let us recall that
one of our major assumption is that these sizes are small
compared to the CCC inner and outer radii.
Both parameters are swept over the range $[0.5, 2.5]$mm,
while keeping the other parameters constant.
Before analyzing the results, it is important to stress the following fact:
even if the air gap or the thickness do not appear explicitly
as a parameter in our model~\eqref{eq:folded:glob} and~\eqref{eq:folded:a},
they impact the considered sequence of radii $R_i$.
Thus, as the superconductor thickness or the air gap changes,
the solution given by~\eqref{eq:folded:glob} and~\eqref{eq:folded:a}
will also change.

\begin{figure}[h!]
  \centering
  \includegraphics{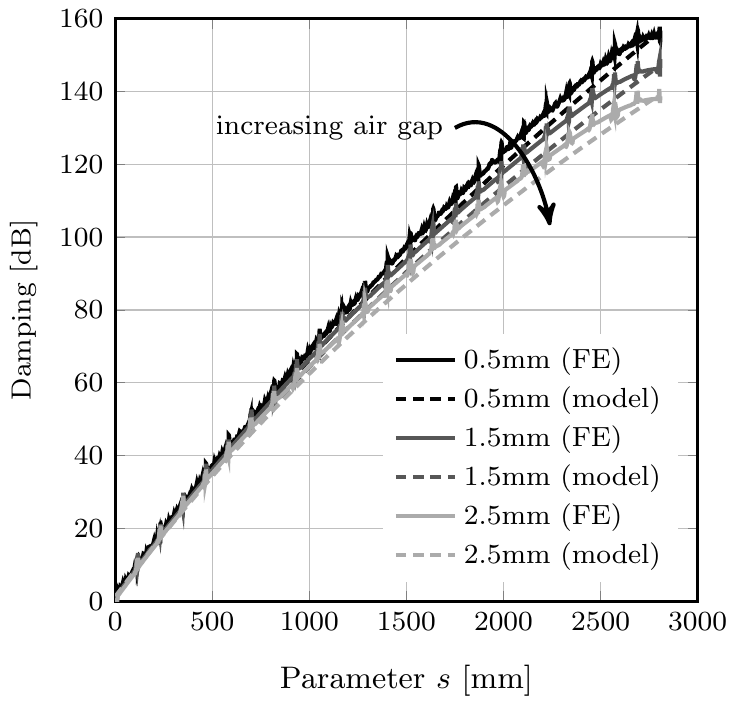}
  \caption{Influence of the air gap (inner variant).}
  \label{fig:folded:g:dw}
\end{figure}

\begin{figure}[h!]
  \centering
  \includegraphics{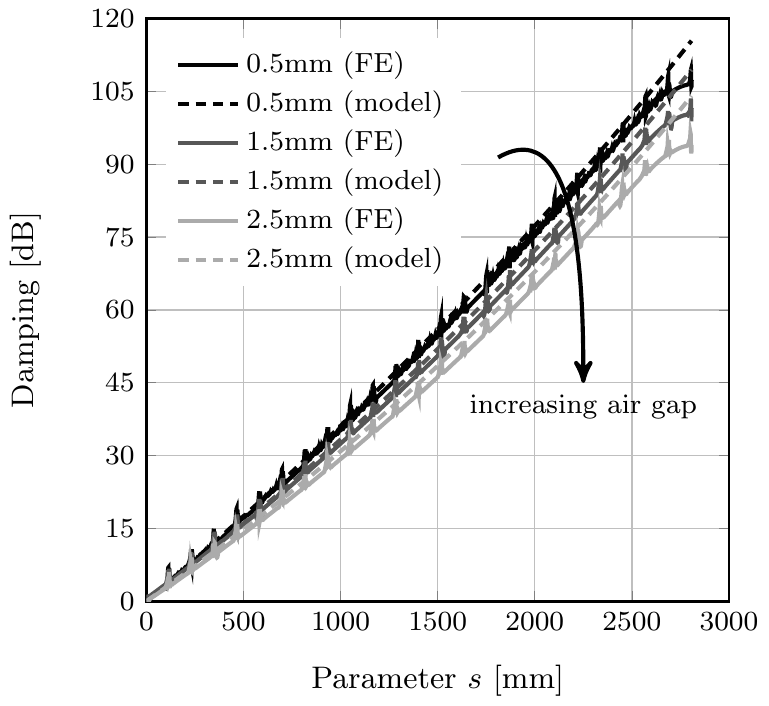}
  \caption{Influence of the air gap (outer variant).}
  \label{fig:folded:g:up}
\end{figure}

\begin{figure}[h!]
  \centering
  \includegraphics{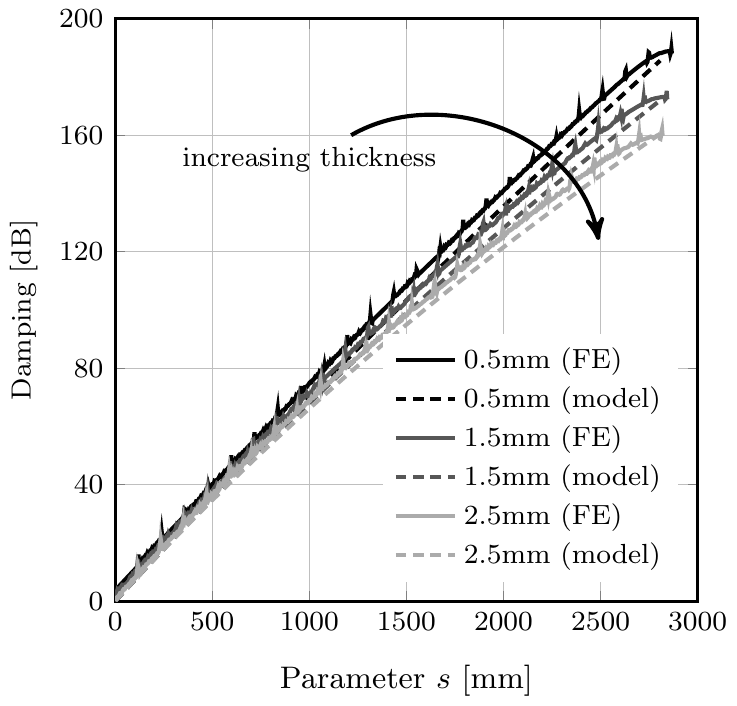}
  \caption{Influence of the thickness
    of the superconducting material (inner variant).}
  \label{fig:folded:t:dw}
\end{figure}

\begin{figure}[h!]
  \centering
  \includegraphics{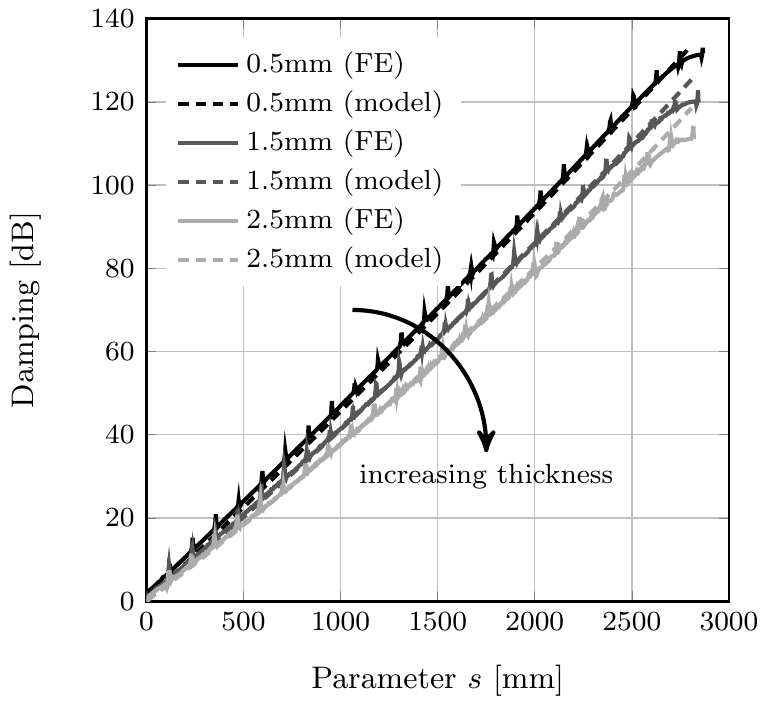}
  \caption{Influence of the thickness
    of the superconducting material (outer variant).}
  \label{fig:folded:t:up}
\end{figure}

The computed results of the damping behavior with respect to the air gap size
and the material thickness are shown in Figures~\ref{fig:folded:g:dw},
\ref{fig:folded:g:up}, \ref{fig:folded:t:dw}
and~\ref{fig:folded:t:up}.
Notice, an increase of the air gap size
corresponds to a decrease of the shielding performance,
both for the inner and outer variants.
Concerning the thickness of the superconducting material,
the same behavior as for the air gap size can be observed.
Finally, it worth noticing that for each simulation presented in
Figures~\ref{fig:folded:g:dw}, \ref{fig:folded:g:up},
\ref{fig:folded:t:dw} and~\ref{fig:folded:t:up},
the maximum relative deviation between the analytical model and the FE simulation did not exceed~$10\%$.

\section{Review of the ring shield}
\label{sec:ring}
\begin{figure}[ht]
  \centering
  \includegraphics{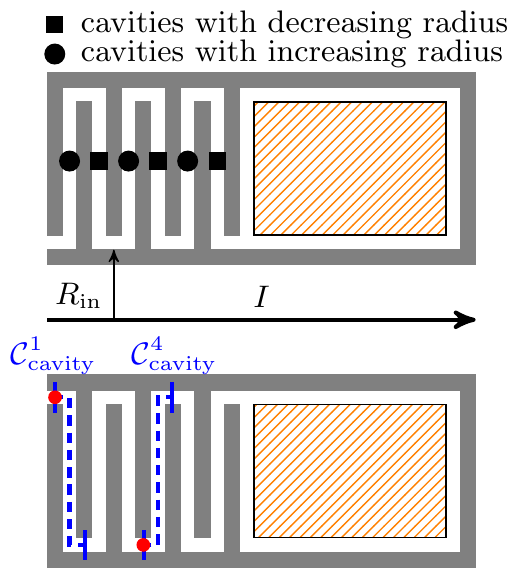}
  \caption{Ring shield configuration (axial cut, two dimensional schematic);
    the symbol~\origin~indicates the origin of a curve
    (\ie{}~the location of its parametric coordinate $s_\text{min}$).}
  \label{fig:ring:2}
\end{figure}

\begin{figure}[ht]
  \centering
  \includegraphics[width=4.75cm]{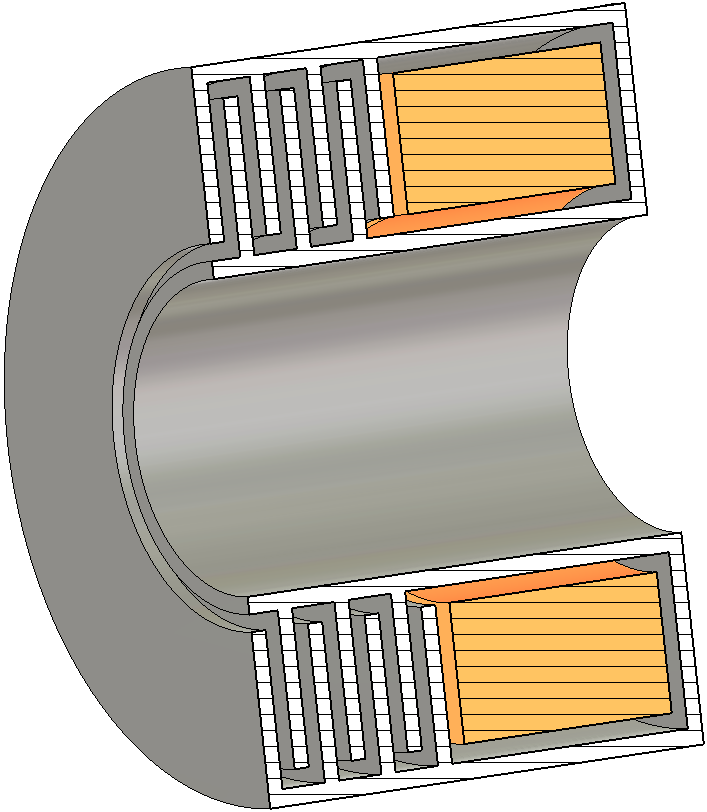}
  \caption{Ring shield configuration (axial cut, three dimensional CAD view).}
  \label{fig:ring:3}
\end{figure}

Ring shield CCCs are commonly applied in diagnostic devices of particle beams in
accelerator facilities~\cite{Peters1998, Tanabe1999, Geithner2012, Fernandes2017}.
This shielding configuration exhibits an interleaved comb-like meander structure
along the axial direction, as depicted in Figures~\ref{fig:ring:2}
and~\ref{fig:ring:3}.
This geometry is constructed by successively stacking superconducting rings
with different inner and outer radii~\cite{DeGersem2016}
(hence the name of ring shield).
The ring stack gives rise to two families of cavities,
as suggested by Figure~\ref{fig:ring:2}:
cavities with increasing and decreasing radius.

The local attenuation of cavities with alternating, increasing or decreasing, radius
writes, for all $s~\in~[\ell_{i-1},~\ell_i]$, as follows~\cite{Grohmann1976b}:
\begin{subequations}
  \label{eq:ring:loc}
  \begin{align}
    \delta_{\mathcal{C}^i_\text{cavity}}(s)
    & = \displaystyle\left(\frac{s-\ell_{i-1}}{R_\text{in}}+1\right)^2
    & \forall i~\text{odd\footnotemark}\\
    \delta_{\mathcal{C}^i_\text{cavity}}(s)
    & = 1
    & \forall i~\text{even\footnotemark},
  \end{align}
\end{subequations}%
\addtocounter{footnote}{-1}\footnotetext{Increasing radius cavities.}%
\stepcounter{footnote}\footnotetext{Decreasing radius cavities.}%
where the curves upon which these local dampings are defined are depicted
in Figure~\ref{fig:ring:2}, and
$\ell_i$ is defined in the same way as in the folded coaxial case
and $R_\text{in}$ is the inner radius of the CCC.
Further details can be found in~\cite{Grohmann1976b}.
From Equation~\eqref{eq:ring:loc},
it can be directly noticed that only the cavities with an increasing radius
are contributing to the local damping,
the only goal of a decreasing radius cavity being to connect two increasing cavities.
While this behavior can be formally derived
from the London-Amp\`ere equations~\cite{Grohmann1976b},
a geometrically driven approach can be found in~\cite{DeGersem2016}.
Finally, the global damping profile is derived following
the same strategy as in the previous section, leading to:
\begin{subequations}
  \label{eq:ring:glob}
  \begin{align}
    d_{\mathcal{C}^i_\text{cavity}}(s)
    & = \alpha_i\left(\frac{s-\ell_{i-1}}{R_\text{in}}+1\right)^2
    & \forall i~\text{odd},\\
    d_{\mathcal{C}^i_\text{cavity}}(s)
    & = \alpha_i
    & \forall i~\text{even},
  \end{align}
\end{subequations}
for all $ s\in[\ell_{i-1}, \ell_i]$, with $\alpha_1=1$ and
\begin{equation}
  \alpha_{i+1} = \alpha_i \times \left\{
    \begin{array}{l@{~}r}
      \displaystyle\left(\frac{\ell_i-\ell_{i-1}}{R_\text{in}}+1\right)^2
      & \,\,\,\forall i~\text{odd},
      \\
      1
      & \,\,\,\forall i~\text{even}.
    \end{array}
  \right.
  \label{eq:ring:a}
\end{equation}

This last relation is proved by exploiting~\eqref{eq:folded:a:def},
and by definition of $\mathcal{C}^1_\text{cavity}$,
the equality $\alpha_1=1$ comes directly,
since $B_\text{in}$ equals $B(s_\text{min})$ on this curve.
Furthermore, by combining the definitions~\eqref{eq:ring:glob}
and~\eqref{eq:folded:a:def}, we can write\footnote{In what follows,
  the ``$\forall i~\text{odd}$/$\forall i~\text{even}$'' choice
  is omitted for conciseness reasons and
  the convention of equation~\eqref{eq:ring:a} is followed.
  Furthermore, let us mention that the following equations are valid
  $\forall s\in[\ell_{i-1}, \ell_i]$.}:
\begin{align}
  \label{eq:ring:induction:i}
  d_{\mathcal{C}_\text{cavity}^i}(s)
  & = d_{\mathcal{C}_\text{cavity}^{i}}(\ell_{i-1})\times
    \left\{
    \begin{array}{l@{~}r}
      \displaystyle\left(\frac{s-\ell_{i-1}}{R_\text{in}}+1\right)^2,
      \\
      1.
    \end{array}
  \right.
\end{align}
We can then evaluate~\eqref{eq:ring:induction:i} at $s=\ell_i$,
and by exploiting the continuity condition~\eqref{eq:folded:continuity} we get:
\begin{align}
  \label{eq:ring:induction:i+1}
  d_{\mathcal{C}_\text{cavity}^i}(\ell_i)
  & = d_{\mathcal{C}_\text{cavity}^{i+1}}(\ell_i),\nonumber\\
  & = d_{\mathcal{C}_\text{cavity}^i}(\ell_{i-1})\times
    \left\{
    \begin{array}{l@{~}r}
      \displaystyle\left(\frac{\ell_i-\ell_{i-1}}{R_\text{in}}+1\right)^2,
      \\
      1.
    \end{array}
    \right.
\end{align}
Then, by combining \eqref{eq:folded:a:def} and~\eqref{eq:ring:induction:i+1},
we can conclude the proof:
\begin{equation}
  \label{eq:ring:induction:proof}
  \alpha_{i+1} = \alpha_i \times \left\{
    \begin{array}{l@{~}r}
      \displaystyle\left(\frac{\ell_i-\ell_{i-1}}{R_\text{in}}+1\right)^2
      & \,\,\,\forall i~\text{odd},
      \\
      1
      & \,\,\,\forall i~\text{even}.
    \end{array}
  \right.
\end{equation}

From these last results,
it worth noticing that the total damping exhibited by
a shield composed of $2N$ rings writes:
\begin{equation*}
  \begin{array}{ll}
    & d_{\mathcal{C}_\text{cavity}^{2N}}(\ell_{2N})\\
  = & \alpha_{2N}
      \times
      \delta_{\mathcal{C}_\text{cavity}^{2N}}(\ell_{2N}),\\
  = & \alpha_{2N}
      \times
      1,
    \\
  = & \alpha_{2N-1}
      \times
      \left(\frac{\ell_{2N-1}-\ell_{2N-2}}{R_\text{in}}+1\right)^2,
    \\
  = & \alpha_{2N-2}
      \times
      1
      \times
      \left(\frac{\ell_{2N-1}-\ell_{2N-2}}{R_\text{in}}+1\right)^2,
    \\
  = & \alpha_{2N-3}
      \times
      \left(\frac{\ell_{2N-3}-\ell_{2N-4}}{R_\text{in}}+1\right)^2
    \\
    & \phantom{\alpha_{2N-3}}
      \times\left(\frac{\ell_{2N-1}-\ell_{2N-2}}{R_\text{in}}+1\right)^2,
    \\
  = & \dots,
    \\
  = & \alpha_{2N-(2N-1)}
      \times
      \left(\frac{\ell_{2N-(2N-1)}-\ell_{2N-(2N-1)-1}}{R_\text{in}}+1\right)^2
    \\
    & \phantom{\alpha_{2N-(2N-1)}}
      \times
      \dots
    \\
    & \phantom{\alpha_{2N-(2N-1)}}
      \times
      \left(\frac{\ell_{2N-3}-\ell_{2N-4}}{R_\text{in}}+1\right)^2,
  \end{array}
\end{equation*}
which leads to
\begin{equation}
  d_{\mathcal{C}_\text{cavity}^{2N}}(\ell_{2N}) =
  \prod_{i=1}^N\left(\frac{\ell_{2i-1}-\ell_{2i-2}}{R_\text{in}}+1\right)^2,
\end{equation}
and is, as expected, nothing but the product of the damping introduced
by each pair of increasing-decreasing rings.

\section{Volumetric performance comparison}
\label{sec:perf}
By comparing the damping profiles of a ring~\eqref{eq:ring:loc}
and a folded coaxial shield~\eqref{eq:folded:loc},
the main differences between the two topologies appear clearly,
and are summarized in Table~\ref{tab:comparison}.
The first motivation to choose one topology over the other
is related to space constraints.
Depending on the available space, one may be interested in building
a CCC expanding in the radial or the axial direction.
On the other hand, if space is not a constraint,
one is interested in building the lightest shield
for a given total damping,
\ie{} a shield exhibiting the smallest volume.
In this case, the exponential damping profile of the folded coaxial shield
seems the most attractive.
However, the situation is unfortunately not that simple.
\begin{table*}[ht]
  \centering
  \begin{tabular}{lll}
    \hline
    \hline
    Parameter        & Folded coaxial & Ring\\
    \cmidrule(lr){1-1}\cmidrule(lr){2-2}\cmidrule(lr){3-3}
    Damping profile  & Exponential    & Quadratic (increasing ring)\\
    of a layer       &                & No damping (decreasing ring)\\
    \\[-0.5em]
    Efficiency of each & Decreases (inner variant) & Constant\\
    additional layer   & Increases (outer variant)\\
    \\[-0.5em]
    Stacking direction & Radial         & Axial\\
    \hline
    \hline
  \end{tabular}
  \caption{Comparison between folded coaxial and ring shields.}
  \label{tab:comparison}
\end{table*}

\subsection{Single layer case}
\begin{figure}[ht]
  \centering
  \includegraphics{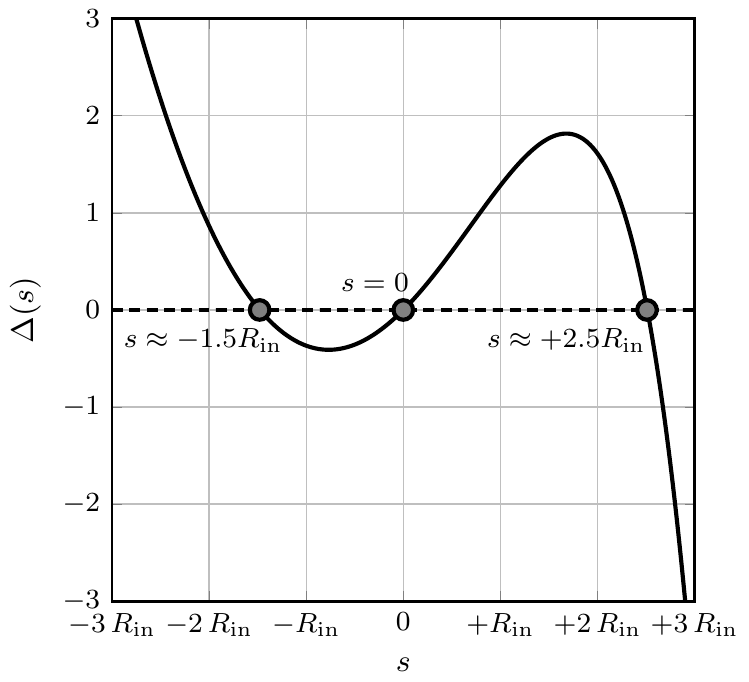}
  \caption{Graph of $\Delta(s)$.}
  \label{fig:comparison:delta}
\end{figure}

Let us consider a single increasing radius cavity and a coaxial shell.
The local dampings are given by~\eqref{eq:ring:loc} and~\eqref{eq:folded:loc} respectively.
Furthermore, we have that
$i=1$, $\ell_{i-1} = 0$ and $R_i = R_\text{in}+g \approx R_\text{in}$,
since we assumed the air gap $g$ to be negligible with respect to $R_\text{in}$.
In this case, we have:
\begin{align}
  \Delta(s)
  & = \delta_{\mathcal{C}_\text{cavity}^1}(s)-
      \delta_{\mathcal{C}_\text{folded}^1}(s),
  \nonumber\\
  & = \left(\frac{s}{R_\text{in}}+1\right)^2 -
      \exp\left(\frac{s}{R_\text{in}}\right).
\end{align}
The real-valued roots of the function $\Delta(s)$
are found
using the symbolic solver of Wolfram Mathematica~\cite{Mathematica2016},
and are given by:
\begin{equation*}
  \begin{array}{r@{~}l}
    \displaystyle\frac{s}{R_\text{in}}\in
      \Bigg\{
    & \overbrace{-1-2W_0\left(\frac{1}{2\sqrt{e}}\right)}^{\approx -1.5}, \\
    & 0, \\
    & \underbrace{-1-2W_{-1}\left(-\frac{1}{2\sqrt{e}}\right)}_{\approx +2.5}
      \Bigg\},
  \end{array}
\end{equation*}
where $W_0$ and $W_{-1}$ are the two real branches of the
Lambert $W$ function~\cite{Corless1996}.
Furthermore, by analyzing the graph of $\Delta(s)$,
depicted in Figure~\ref{fig:comparison:delta},
it is clear that this function is:
\begin{itemize}
\item positive for $s\in]-\infty,-1.5R_\text{in}[$
  and \\$s\in]0, 2.5R_\text{in}[$;
\item negative for $s\in]-1.5R_\text{in}, 0[$
  and \\$s~\in~]2.5R_\text{in}, +\infty[$.
\end{itemize}
These results leads us then to the following conclusion:
let an increasing radius ring and a coaxial shell
of the same size $\tilde{\ell}$ be given,
then the following holds
\begin{equation}
  \delta_{\mathcal{C}_\text{folded}^1}(\tilde{\ell}) \geq
  \delta_{\mathcal{C}_\text{cavity}^1}(\tilde{\ell})
  \quad \text{if}\quad
  \tilde{\ell} \geq 2.5R_\text{in}.
  \label{eq:comparison:l}
\end{equation}%
Therefore, despite its exponential damping,
the folded coaxial CCC will not necessarily exhibit a better damping
than its ring alternative, at least for the first ring
or coaxial shell.

\subsection{General case}
For a more realistic analysis of volume and shielding performances,
the following numerical experiment is carried out:
for the three CCC shields presented above (ring, inner folded coaxial and
outer folded coaxial) we compute the damping corresponding to
different axial lengths $L$, outer radius sizes $R_\text{out}$
and numbers of meanders.
Among all these configurations,
the inner radius $R_\text{in}$ and
the core cross section $S_\text{core}$ are held constant,
as reported in Table~\ref{tab:perf:const}.
Then, for each result,
the shield volume is approximated
by using the equations given in Table~\ref{tab:perf:V:def}.
\begin{table}[ht]
  \centering
  \begin{tabular}{cc}
    \hline
    \hline
    \multicolumn{1}{c}{Constrained parameter} &
    \multicolumn{1}{c}{Constraint value}\\
    \cmidrule(lr){1-1}\cmidrule(lr){2-2}
    Inner radius ($R_\text{in}$) & $120$mm\\
    Core cross section ($S_\text{core}$) & $60$cm\up{2}\\
    \hline
    \hline
  \end{tabular}
  \caption{Constraints for the performance comparison.}
  \label{tab:perf:const}
\end{table}
\begin{table*}[ht]
  \centering
  \begin{tabular}{llll}
    \hline
    \hline
    & & \multicolumn{2}{c}{Folded coaxial shield CCC}\\
    \cmidrule(lr){3-4}
    \multicolumn{1}{c}{Volume} &
    \multicolumn{1}{c}{Ring shield CCC} &
    \multicolumn{1}{c}{Inner variant} &
    \multicolumn{1}{c}{Outer variant}\\
    \cmidrule(lr){1-1}\cmidrule(lr){2-2}\cmidrule(lr){3-3}\cmidrule(lr){4-4}
    Total ($V_T$)   & $\pi\,R_\text{out}^2\,L$
                    & $\pi\,R_\text{out}^2\,L$
                    & $\pi\,R_\text{out}^2\,L$\\
    Interior($V_I$) & $\pi\,R_\text{in}^2\,L$
                    & $\pi\,R_\text{in}^2\,L$
                    & $\pi\,R_\text{in}^2\,L$\\
    Core ($V_C$)    & $\pi\,(R_\text{out}^2-R_\text{in}^2)\,L_C$
                    & $V_T-\pi\,(R_\text{out}-H_C)^2\,L$
                    & $\pi\,(R_\text{in}+H_C)^2\,L - V_I$\\
    Shield ($V_S$)  & $\pi\,(R_\text{out}^2-R_\text{in}^2)\,L_S$
                    & $V_T-V_I-V_C$
                    & $V_T-V_I-V_C$\\
    \\
    \multicolumn{1}{c}{Length}\\
    \cmidrule(lr){1-1}
    Core ($L_C$) & $S_\text{core}/H_C$
          & $L$
          & $L$\\
    Shield ($L_S$) & $L-L_C$
          & $L$
          & $L$\\
    \\
    \multicolumn{1}{c}{Height}\\
    \cmidrule(lr){1-1}
    Core ($H_C$) & $R_\text{out}-R_\text{in}$
          & $S_\text{core}/L$
          & $S_\text{core}/L$\\
    \hline
    \hline
  \end{tabular}
  \caption{Volumes definitions.}
  \label{tab:perf:V:def}
\end{table*}

\begin{figure*}[p]
  \centering
  \begin{subfigure}{0.49\linewidth}
    \centering
    \includegraphics{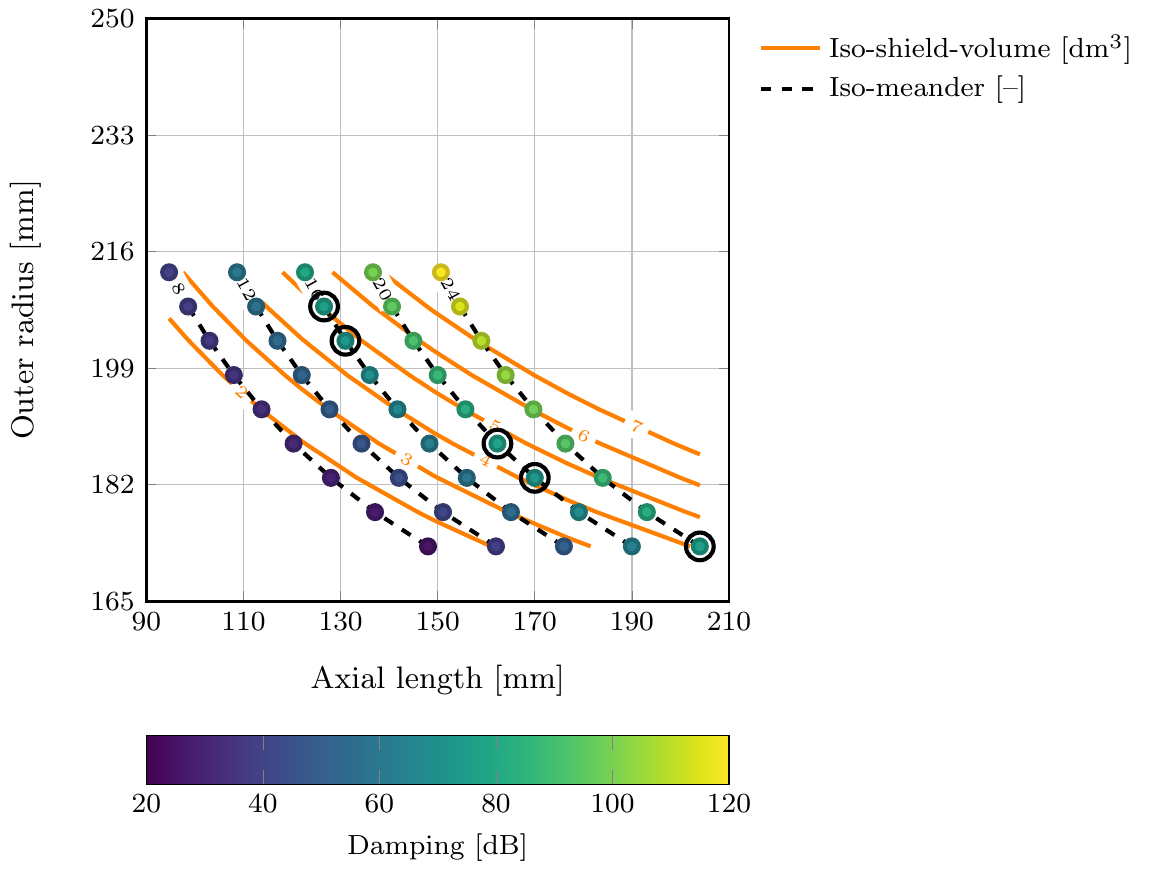}
    \caption{Ring shield.}
  \end{subfigure}

  \begin{subfigure}{0.49\linewidth}
    \centering
    \includegraphics{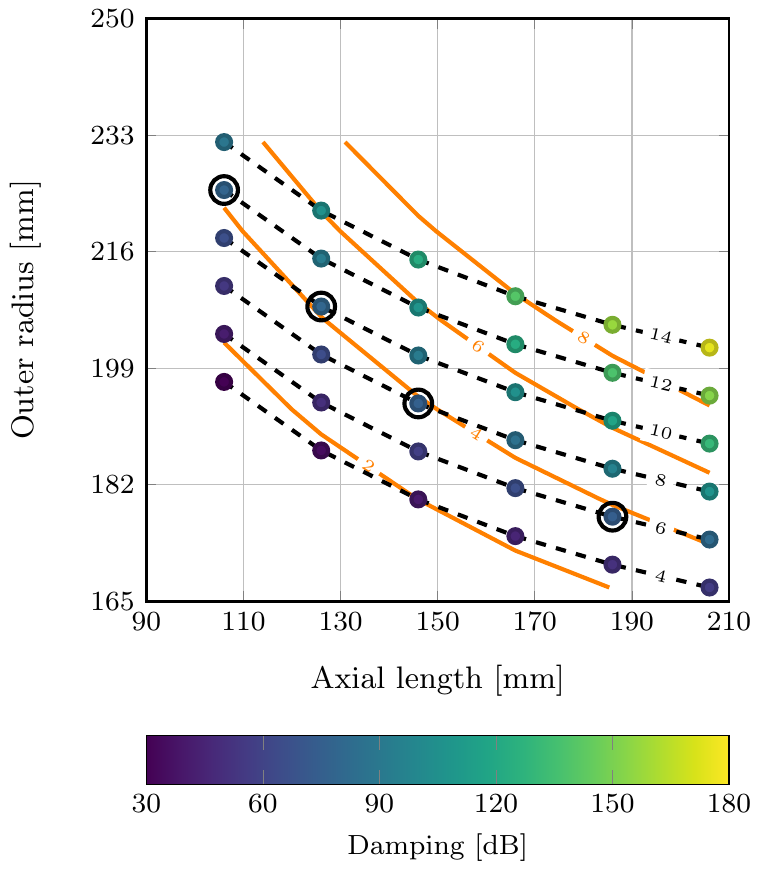}
    \caption{Folded coaxial shield (inner variant).}
  \end{subfigure}
  \begin{subfigure}{0.49\linewidth}
    \centering
    \includegraphics{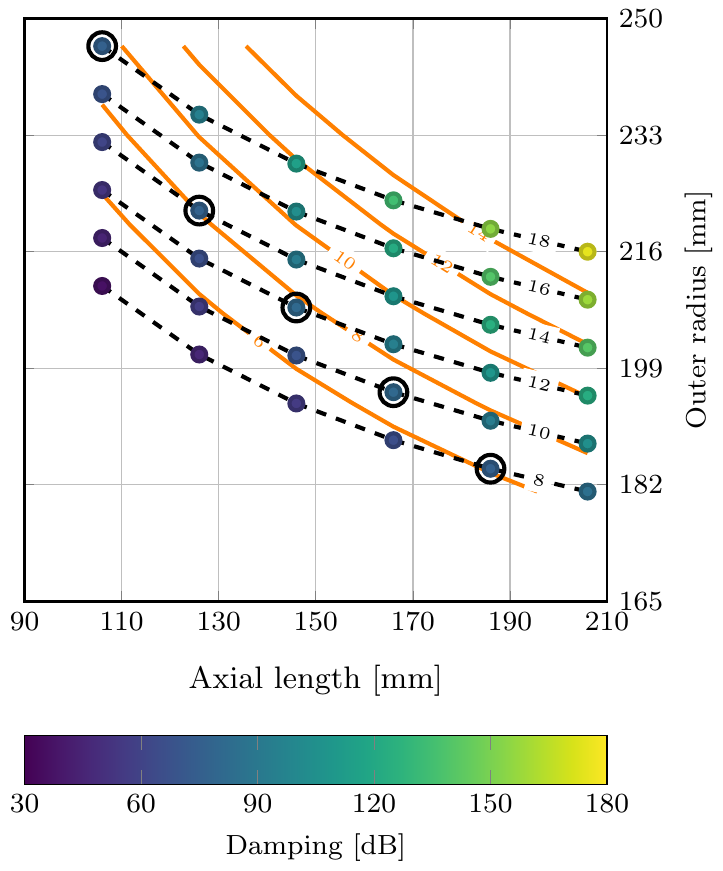}
    \caption{Folded coaxial shield (outer variant).}
  \end{subfigure}
  \caption{Shield volumetric performance for different configurations, where
    the encircled configurations exhibit a damping
    in the range $75$dB $\pm5$dB
    and the fixed parameters are set to $R_\text{in}=120$mm
    and $S_\text{core}=60$cm\up{2}.}
  \label{fig:perf:shield}
\end{figure*}

The computed results are depicted in Figure~\ref{fig:perf:shield}.
The plots are organized as follows.
Each computed configuration is depicted as a colored dot
in the axial length -- outer radius plane.
The color of this dot corresponds to the damping exhibited
by the considered shield.
Among all these possibilities, some are sharing the same number of meanders,
and are thus connected by dashed lines, further called iso-meander lines.
Moreover, since the shield volume is only a function of the axial length
and the outer radius (for a fixed core cross section and inner radius),
it makes sens to additionally represent the iso-shield-volume lines on the plots.

With all these information in hand, the optimization process becomes rather easy.
Since we can directly access the axial length and the outer radius,
we can easily reject or accept a configuration with respect to space constraints.
In our experiment, we restricted ourselves to the arbitrary range:
$L\in[90, 210]$mm and $R_\text{out}\in[165, 250]$mm.
Once this admissible range defined, we can select the configurations sharing
the same damping.
Thanks to the iso-meander lines,
we can directly retrieve the corresponding number of meanders.
It is worth stressing that in the ring topology,
the axial length depends on the number of meanders and the outer radius
is a free parameter,
while the situation is reverted for the folded coaxial cases.
In our numerical setup, we chose to consider dampings in the range
$75$dB $\pm5$dB.
Then, we can search on which iso-shield-volume lines
the selected candidates are lying.
Finally,
the configuration leading to the minimal shield volume can be selected,
as shown in Table~\ref{tab:perf:best}.

\begin{table*}[ht]
  \centering
  \begin{tabular}{ccccccc}
    \hline
    \hline
    & \multicolumn{2}{c}{Volume}\\
    \cmidrule(lr){2-3}
    Type & Shield & Core & Damping & Outer radius & Axial length & Shell/Ring\\
    \cmidrule(lr){1-1}\cmidrule(lr){2-2}\cmidrule(lr){3-3}
    \cmidrule(lr){4-4}\cmidrule(lr){5-5}\cmidrule(lr){6-6}\cmidrule(lr){7-7}
    Ring  & $4.1$dm\up{3} & $5.9$dm\up{3} & $76$dB & $173$mm & $204$mm & $24$\\
    Inner & $3.7$dm\up{3} & $6.3$dm\up{3} & $74$dB & $177$mm & $186$mm & \phantom{0}$6$\\
    Outer & $6.1$dm\up{3} & $5.3$dm\up{3} & $75$dB & $184$mm & $186$mm & \phantom{0}$8$\\
    \hline
    \hline
  \end{tabular}
  \caption{Shields with minimal volumes,
    where \textit{inner} and \textit{outer} designate
    the inner and outer variants of the folded coaxial shield
    (fixed parameters $R_\text{in}=120$mm and $S_\text{core}=60$cm\up{2}).}
  \label{tab:perf:best}
\end{table*}

By analyzing the damping and volume characteristics reflected in Figure~\ref{fig:perf:shield},
the following conclusion can be drawn.
In the case of the ring configuration, the shield volume is minimized
by favoring small outer radii and a large number of meanders.
On the other hand, in the case of the folded coaxial topology,
the shield volume is minimized when a small amount of long meanders is used.
This behavior is easily explained,
since the shield volume depends quadratically on the shield outer radius
and linearly on the shield length.
Therefore, a reduction of the outer radius is more effective in the ring configuration,
even if this reduction calls for an increase in the number of meander,
and thus an increase in the shield length.
The same reasoning can be applied for the folded coaxial case:
a decrease of the outer radii calls for a reduction in the number of meanders,
which must be compensated by increasing the length of one meander.

Furthermore, by exploiting the data presented in Table~\ref{tab:perf:best},
it becomes clear that among the different shield topologies,
the inner variant of the folded coaxial approach leads to the lightest
(\ie{} with the smallest volume) shield.
However, the weight of the total CCC is also determined by its magnetic core.
If we now analyze its occupied volume, the outer counterpart becomes more
interesting.
It is then worth mentioning that
if the densities of the materials used are known,
the presented optimization can be improved:
instead of using iso-shield-volume (or iso-core-volume) lines,
it is more interesting to directly use iso-mass lines,
computed by weighting the core and shield volumes by their respective densities.

\section{Conclusion}
\label{sec:conclusion}
In this paper,
we analyzed the performance of a folded superconducting coaxial shield
for a cryogenic current comparator.
The damping profile presented by this new shield topology was first estimated
with an analytical model, which was then validated by means of finite element simulations.
In all the carried out simulations,
the relative difference between the two approaches did not exceed~$10\%$.

By analyzing the newly proposed analytical model, we realized that the damping efficiency
of each additional coaxial layer was not constant, as in the ring topology.
Indeed, depending on whether the CCC opening is located at the inner or outer radius,
the damping efficiency of an additional shielding layer decreases or increases,
respectively.
Furthermore, because of the space taken by the detection coil,
the radii of the coaxial shells cannot span the same range.
Therefore, the overall performance of the inner and outer variants cannot be identical.

Finally, this work compared the ring topology
with the two variants of the folded coaxial shield.
At this point, different shield configurations were tested,
in which the inner radius and the core cross section have been held constant
(in particular, these parameters were fixed to $120$mm and $60$cm\up{2}).
Among all the considered configuration, only those leading to a total damping
in the range $75$dB $\pm5$dB were selected.
This numerical experiment led us to the following conclusions:
\begin{enumerate*}[label = \itshape\roman*\upshape)]
\item in the case of a ring topology,
  the shield volume is minimized by favoring small outer radii
  and a large number of meanders;

\item in the case of a folded coaxial topology,
  the shield volume is minimized by favoring a small amount of long meanders.
\end{enumerate*}

\section*{Acknowledgment}
This research is funded by
the German Bundesministerium f\"ur Bildung und Forschung
as the project BMBF-05P15RDRBB
``Ultra-Sensitive Strahlstrommessung f\"ur zuk\"unftige Beschleunigeranlagen''.
Finally, the authors would like to express their gratitude to
the referee for his constructive comments.

\bibliographystyle{elsarticle-num}
\bibliography{biblio}

\begin{thebibliography}{10}
\expandafter\ifx\csname url\endcsname\relax
  \def\url#1{\texttt{#1}}\fi
\expandafter\ifx\csname urlprefix\endcsname\relax\def\urlprefix{URL }\fi
\expandafter\ifx\csname href\endcsname\relax
  \def\href#1#2{#2} \def\path#1{#1}\fi

\bibitem{Harvey1972}
I.~K. Harvey, A precise low temperature dc ratio transformer, Review of
  Scientific Instruments 43~(11) (1972) 1626--1629.
\newblock \href {https://doi.org/10.1063/1.1685508}
  {\path{doi:10.1063/1.1685508}}.

\bibitem{Williams2011}
J.~M. Williams, Cryogenic current comparators and their application to
  electrical metrology, IET Science, Measurement and Technology 5~(6) (2011)
  211--224.
\newblock \href {https://doi.org/10.1049/iet-smt.2010.0170}
  {\path{doi:10.1049/iet-smt.2010.0170}}.

\bibitem{Clarke2004}
J.~Clarke, A.~I. Braginski (Eds.), The {S}{Q}{U}{I}{D} Handbook: Fundamentals
  and Technology of {S}{Q}{U}{I}{D}s and {S}{Q}{U}{I}{D} Systems, Wiley-VCH,
  Weinheim, 2004.
\newblock \href {https://doi.org/10.1002/3527603646}
  {\path{doi:10.1002/3527603646}}.

\bibitem{Sullivan1974}
D.~B. Sullivan, R.~F. Dziuba, Low temperature direct current comparators,
  Review of Scientific Instruments 45~(4) (1974) 517--519.
\newblock \href {https://doi.org/10.1063/1.1686674}
  {\path{doi:10.1063/1.1686674}}.

\bibitem{Grohmann1974}
K.~Grohmann, H.~D. Hahlbohm, H.~L\"ubbig, H.~Ramin, Ironless cryogenic current
  comparators for {AC} and {DC} applications, IEEE Transactions on
  Instrumentation and Measurement 23~(4) (1974) 261--263.
\newblock \href {https://doi.org/10.1109/TIM.1974.4314287}
  {\path{doi:10.1109/TIM.1974.4314287}}.

\bibitem{Grohmann1976a}
K.~Grohmann, H.~D. Hahlbohm, D.~Hechtfischer, H.~L\"ubbig, Field attenuation as
  the underlying principle of cryo current comparators, Cryogenics 16~(7)
  (1976) 423--429.
\newblock \href {https://doi.org/10.1016/0011-2275(76)90056-4}
  {\path{doi:10.1016/0011-2275(76)90056-4}}.

\bibitem{Grohmann1976b}
K.~Grohmann, H.~D. Hahlbohm, D.~Hechtfischer, H.~L\"ubbig, Field attenuation as
  the underlying principle of cryo-current comparators 2. {R}ing cavity
  elements, Cryogenics 16~(10) (1976) 601--605.
\newblock \href {https://doi.org/10.1016/0011-2275(76)90192-2}
  {\path{doi:10.1016/0011-2275(76)90192-2}}.

\bibitem{Seppa1990}
H.~Seppa, The ratio error of the overlapped-tube cryogenic current comparator,
  IEEE Transactions on Instrumentation and Measurement 39~(5) (1990) 689--697.
\newblock \href {https://doi.org/10.1109/19.58609}
  {\path{doi:10.1109/19.58609}}.

\bibitem{Peters1998}
A.~Peters, W.~Vodel, H.~Koch, R.~Neubert, H.~Reeg, C.~H. Schroeder, A cryogenic
  current comparator for the absolute measurement of {nA} beams, AIP Conference
  Proceedings 451~(1) (1998) 163--180.
\newblock \href {https://doi.org/10.1063/1.56997} {\path{doi:10.1063/1.56997}}.

\bibitem{Tanabe1999}
T.~Tanabe, K.~Chida, K.~Shinada, A cryogenic current-measuring device with
  nano-ampere resolution at the storage ring {TARN} {II}, Nuclear Instruments
  and Methods in Physics Research Section A: Accelerators, Spectrometers,
  Detectors and Associated Equipment 427~(3) (1999) 455--464.
\newblock \href {https://doi.org/10.1016/S0168-9002(99)00058-3}
  {\path{doi:10.1016/S0168-9002(99)00058-3}}.

\bibitem{Geithner2012}
R.~Geithner, W.~Vodel, R.~Neubert, P.~Seidel, F.~Kurian, H.~Reeg,
  M.~Schwickert, An improved cryogenic current comparator for {FAIR}, in: 3rd
  International Particle Accelerator Conference IPAC'12, 2012, pp. 822--824.

\bibitem{Fernandes2017}
M.~Fernandes, R.~Geithner, J.~Golm, R.~Neubert, M.~Schwickert, T.~St\"ohlker,
  J.~Tan, C.~P. Welsch, Non-perturbative measurement of low-intensity charged
  particle beams, Superconductor Science and Technology 30~(1) (2017) 015001.
\newblock \href {https://doi.org/10.1088/0953-2048/30/1/015001}
  {\path{doi:10.1088/0953-2048/30/1/015001}}.

\bibitem{Cst2015}
{Computer Simulation Technology AG}, {CST EM STUDIO\up{\copyright}},
  \url{www.cst.com} (2015).

\bibitem{DeGersem2016}
H.~De~Gersem, N.~Marsic, W.~F.~O. M\"uller, F.~Kurian, T.~Sieber,
  M.~Schwickert, Finite-element simulation of the performance of a
  superconducting meander structure shielding for a cryogenic current
  comparator, Nuclear Instruments and Methods in Physics Research Section A:
  Accelerators, Spectrometers, Detectors and Associated Equipment 840 (2016)
  77--86.
\newblock \href {https://doi.org/10.1016/j.nima.2016.10.003}
  {\path{doi:10.1016/j.nima.2016.10.003}}.

\bibitem{Mathematica2016}
{Wolfram Research, Inc.}, Mathematica, {V}ersion 11.0, {C}hampaign, IL (2016).

\bibitem{Corless1996}
R.~M. Corless, G.~H. Gonnet, D.~E.~G. Hare, D.~J. Jeffrey, D.~E. Knuth, On the
  {L}ambert {W} function, Advances in Computational Mathematics 5~(1) (1996)
  329--359.
\newblock \href {https://doi.org/10.1007/BF02124750}
  {\path{doi:10.1007/BF02124750}}.

\end{thebibliography}
\end{document}